# Twisted Electron Collisions Enhance the Production of Circular Rydberg States

H. S. Parker, S. L. Sims, D. J. Lamphere, and A. L. Harris[*]

Department of Physics, Illinois State University, Normal, IL 61790, USA

## Abstract

Circular Rydberg states offer advantages for quantum information and quantum simulation platforms due to their long lifetimes and strong dipole–dipole interactions. Unfortunately, current techniques for the production of these states remain technically challenging. Here we investigate the ability of twisted electron collisions to produce circular Rydberg states. Twisted electrons carry quantized orbital angular momentum that can be transferred to the electronic state of the atom, potentially providing an efficient means to generate circular Rydberg states. Using a fully quantum mechanical framework, we compute total excitation cross sections for circular Rydberg states of hydrogen, rubidium, and cesium targets using Bessel electron beams. Our models account for the full Bessel-beam structure of the incident electron and incorporate macroscopic target effects to model experimentally-relevant conditions. Our results show that twisted electrons with large opening angles produce significant enhancements in the excitation probability relative to plane-wave electrons, particularly for large opening angles and low energies. We trace this enhancement to contributions from projectiles with large values of orbital angular momentum. These findings demonstrate that twisted-electron excitation may provide a feasible and potentially advantageous pathway for generating circular Rydberg states.

## 1. Introduction

Rydberg atoms have become important players in the quantum computing and information revolution with neutral-atom arrays emerging as one of the leading architectures for quantum computing platforms [1,2]. In these architectures, the Rydberg atoms act as qubits, enabling the formation of logic gates [1,3] performed using optical and microwave fields. These arrays offer large-scale, flexible geometries with relatively long lifetimes and strong interactions [2,4,5]. The large dipole moments and strong long-range interactions present in Rydberg atoms have led to high fidelity and multi-qubit states [4–6]. However, the feasibility of these systems is limited by the lifetime of the Rydberg states [4], which scale as $n^3$. Some improvement has been made through the use of circular Rydberg atoms, which possess the highest possible value of angular momentum. The lifetime of circular Rydberg states scales as $n^5$ [7], and under the right experimental conditions, lifetimes can range from several milliseconds [1,8] to 100 seconds [9,10]. This makes circular Rydberg states attractive candidates for use in neutral atom arrays.

[*]Corresponding author: alharri@ilstu.edu

Aside from quantum computing, Rydberg atoms have also shown their potential usefulness in quantum simulation [1] for the study of many-body physics where neutral atom arrays offer a convenient technique to study spin models [2]. Here, too, circular Rydberg atoms may offer advantages not found in Rydberg arrays with low angular momentum states. For example, Rydberg atoms trapped with Laguerre-Gauss optical beams can serve as sensitive detectors of dipole-dipole interactions [7] with the low autoionization rates and longer lifetimes providing opportunities to study effects over longer time scales [7,11,12].

While progress has been made in the advancement of Rydberg atom arrays and the control of individual atoms, the use of circular Rydberg states for quantum computing is often limited by the technical challenges inherent in their creation. Current methods for generating circular Rydberg states include multiphoton excitation processes [4] that obey the usual dipole selection rules. This makes it challenging to reach high $n$ states due to the large number of photons required. Alternatives include rapid adiabatic passage [13,14], coupling of a low angular momentum Rydberg state to a radio frequency field [15], and application of two-color circularly polarized pulses [16]. One potentially advantageous alternative to Rydberg state creation is the use of optical vortex beams, which have been shown to transfer angular momentum directly to the electronic state of the atom and to enable the relaxation of the standard dipole selection rules [17–19].

Another alternative for the creation of circular Rydberg states is excitation through collisions with twisted electrons (also called electron vortex beams) [20]. As with optical vortex beams, electron vortex beams can carry quantized orbital angular momentum (OAM) that can be transferred to both the electronic and the center of mass motion of a target atom [21]. Previous calculations for excitation of one-electron atoms by vortex beams have shown that dipole forbidden transitions can occur with off-axis collisions [22] and that scattering amplitudes are strongly influenced by the amount of orbital angular momentum transferred [23]. These features, combined with the experimental availability of twisted electron beams with high values of OAM [24], open the door to the possibility of generating circular Rydberg states through twisted electron excitation.

We present here excitation cross sections to circular Rydberg states for collisions of twisted electrons with hydrogen, rubidium, and cesium atoms. We show that twisted electrons with large opening angle provide an enhancement in excitation probability over their plane wave counterparts, particularly at low energy. This enhancement was traced to contributions from projectiles with large values of OAM. Our results demonstrate the feasibility of using twisted electron collisions for the production of circular Rydberg states. In section 2, we detail the theoretical methods used to calculate the cross sections. Section 3 includes our results and discussion. Unless otherwise noted, atomic units are used throughout.

## 2. Theory

To examine the feasibility of using twisted electrons to generate circular Rydberg states, we calculate twisted electron total excitation cross sections for Bessel projectiles. The total cross section for a plane wave projectile is given by [25]

$$\sigma(E) = \frac{k_f}{k_i} \int |f(\theta, \varphi)|^2 d\Omega, \tag{1}$$

where $E$ is the incident projectile energy, $\vec{k}_i$ and $\vec{k}_f$ are the incident and scattered projectile momenta, and $f(\theta, \varphi)$ is the scattering amplitude for the projectile scattering with polar angle $\theta$ and azimuthal angle $\varphi$.

In the Born approximation, the scattering amplitude is given by

$$f(\theta, \varphi) = N < \chi_{\vec{k}_f} \Phi_f | V_i | \chi_{\vec{k}_i} \Phi_i >, \tag{2}$$

where $N$ is a normalization constant ($-(2\pi)^2$ for plane waves and $-(2\pi)^{3/2}$ for twisted electrons), $\chi_{\vec{k}_i}$ and $\chi_{\vec{k}_f}$ are the incident and scattered projectile wave functions, $\Phi_i$ and $\Phi_f$ are the initial and final bound state wave functions, and $V_i$ is the perturbation potential given by the Coulomb interaction between the projectile and target atom. In the position representation, the transition matrix can be written as an integral over all coordinate space with the coordinate system chosen such that the target nucleus is located at the origin and $\vec{r}_1$ and $\vec{r}_2$ are the position vectors of the projectile and bound electron. The z-axis is chosen to be along the projectile propagation axis with the projectile scattering at an angle $\theta$ into the x-z plane. This gives

$$f(\theta, \varphi) = N \int d\vec{r}_1 d\vec{r}_2 \, \chi^*_{\vec{k}_f}(\vec{r}_1) \Phi^*_f(\vec{r}_2) \left(-\frac{Z_t}{r_1} + \frac{1}{r_{12}}\right) \chi_{\vec{k}_i}(\vec{r}_1) \Phi_i(\vec{r}_2), \tag{3}$$

where $Z_t$ is the charge of the atomic core.

For the highly excited circular Rydberg states, the quantum defect model is used [26] with the bound states given in spherical coordinates by

$$\Phi(r_2, \theta_2, \varphi_2) = \sqrt{\left(\frac{2}{n'}\right)^3 \frac{(n'-l-1)!}{2n'(n'+l)!}} \, e^{-\frac{r_2}{n'}} \left(\frac{2r_2}{n'}\right)^l L^{2l+1}_{n'-l-1}\left(\frac{2r_2}{n'}\right), \tag{4}$$

where $n' = n - \delta_l$ is the effective principle quantum number, $l$ is the orbital angular momentum quantum number, $m$ is the magnetic quantum number, and $L^{2l+1}_{n'-l-1}\left(\frac{2r_2}{n'}\right)$ is the associated Laguerre

polynomial. The quantum defect $\delta_l$ for large $n$ is given by [27]

$$\delta_l = \delta_0 + \frac{\mu_2}{n^2} + \frac{\mu_4}{n^4}, \tag{5}$$

where $\delta_0$, $\mu_2$, and $\mu_4$ can be written in terms of dipole ($\alpha_d$) and quadrupole ($\alpha_q$) polarizabilities (from [28] and [29]) with

$$\delta_0 = \frac{3}{2}A_4\alpha_d + \frac{35}{2}A_6\alpha_q \tag{6}$$

$$\mu_2 = -\frac{1}{2}\left[l(l+1)A_4\alpha_d + 5(6l^2 + 6l - 5)A_6\alpha_q\right] \tag{7}$$

$$\mu_4 = \frac{3}{2}(l-1)l(l+1)(l+2)A_6\alpha_q \tag{8}$$

$$A_4 = \left[2\left(l - \frac{1}{2}\right)l\left(l + \frac{1}{2}\right)(l+1)\left(l + \frac{3}{2}\right)\right]^{-1} \tag{9}$$

$$A_6 = \frac{A_4}{\left[4\left(l - \frac{3}{2}\right)(l-1)(l+2)\left(l + \frac{5}{2}\right)\right]}. \tag{10}$$

For circular Rydberg states with high $n$ and $l = n - 1$, the energy levels of the bound states are given by

$$E_{n,l} = -\frac{1}{2n'^2}. \tag{11}$$

For lower lying states, including the ground state, the accepted values from NIST [30] are used.

We consider excitation by an incident Bessel projectile, given in cylindrical coordinates by

$$\chi_{\vec{k}_i}(\vec{r}_1) = \frac{e^{il\varphi_1}}{2\pi}J_l(k_{i\perp}\rho_1)e^{ik_{iz}z_1}. \tag{12}$$

The Bessel electron has unique properties not present for a plane wave projectile. First, while it propagates along the z-axis, its momentum vector is not well defined because it lies on a cone centered on the propagation direction. The magnitude of the momentum is fixed, but its azimuthal angle varies from 0 to $2\pi$. This leads to the Bessel projectile momentum having a non-zero transverse component given by

$$k_{i\perp} = k_i \sin \alpha, \qquad (13)$$

where $\alpha$ is the beam's opening angle (half angle of the cone). The longitudinal momentum is given by

$$k_{iz} = k_i \cos \alpha. \qquad (14)$$

Second, the Bessel projectile's density is non-uniform in the transverse direction and has a spatial node and phase singularity along the propagation axis that results from its helical wave fronts [20,31–33]. These features necessitate the specification of an impact parameter $\vec{b}$ to describe the transverse location of the phase singularity relative to the propagation axis. Lastly, the helicity of the wave fronts leads to the Bessel projectile carrying non-zero orbital angular momentum.

In addition to the expression in Eq. (12), the Bessel wave function can also be written as a superposition of tilted plane waves [23]

$$\chi_{\vec{k}_i}(\vec{r}_1) = \frac{(-i)^\lambda}{(2\pi)^2} \int_0^{2\pi} d\phi_{ki} \, e^{i\lambda\phi_{ki}} e^{i\vec{k}_i \cdot (\vec{r}_1 - \vec{b})}, \qquad (15)$$

where $\phi_{ki}$ is the azimuthal angle of the incident projectile momentum and $\lambda$ is the topological charge of the projectile, which indicates the projection of orbital angular momentum onto the propagation axis (often briefly referred to as the OAM of the beam). Eq. (15) clearly shows the uncertainty in incident momentum azimuthal angle, as well as the dependence of the incident wave function on impact parameter.

To calculate the transition amplitude, we assume that the projectile transfers all of its OAM

to the target during the collision. In this case, the scattered electron is represented as a plane wave

$$\chi_{\vec{k}_f}(\vec{r}_1) = \frac{e^{i\vec{k}_f \cdot \vec{r}_1}}{(2\pi)^{3/2}} \tag{16}$$

and the above equations can be combined to calculate the transition amplitude for a twisted Bessel electron with fixed OAM and impact parameter in terms of the plane wave amplitude

$$f^{Bess}(\theta, \varphi) = \frac{(-i)^\lambda}{(2\pi)^{1/2}} \int d\phi_{k_i} f^{PW}(\theta, \varphi). \tag{17}$$

Practically, it is challenging to control the impact parameter of the twisted electron relative to the target and in most charged particle collision experiments, such control is not feasible. Thus, it is necessary for theory to average the cross sections over impact parameter in order to make meaningful predictions or comparisons with experiment. Such an averaging is equivalent to a collision between a twisted electron and a macroscopic collection of randomly distributed target atoms [34]. As detailed in [35–37], this leads to an expression for the cross section as an average over incident momentum azimuthal angles of the plane wave differential cross section

$$\sigma^{Bess} = \frac{k_i}{k_{zi}(2\pi)} \int d\phi_{k_i} \sigma^{PW} d\Omega. \tag{18}$$

Once this average is performed, any explicit dependence on OAM is washed out, leading to a cross section independent of $\lambda$.

### 3. Results

### A. Excitation is more favorable for low energy twisted electrons

The non-zero OAM of electron vortex projectiles makes them potential candidates for generating circular Rydberg states, and the total excitation cross sections provide a measure of the likelihood of generating such states during a collision. We calculated excitation cross sections as a function of projectile energy and vortex opening angle for hydrogen, cesium, and rubidium from their respective ground states.

For a macroscopic target, in which the impact parameter of the projectile relative to the target center is unknown, the cross sections do not explicitly depend on the OAM of the projectile. This averaging has the effect of washing out more pronounced features that may be present due to specific OAM values. Figure 1 shows the total excitation cross sections from the ground state to a circular Rydberg state as a function of energy and opening angle for hydrogen, rubidium, and cesium. For all targets, at small opening angle, the cross sections show a deep minimum at low energy with a broad maximum typically between 100 and 200 eV. This is qualitatively similar to the excitation cross section for a plane wave projectile, which rises sharply for energies just above the excitation threshold, reaches a maximum at an intermediate energy where excitation is most favorable, and then slowly declines at large energies where forward scattering is more favorable. This forward scattering results in a large, transverse momentum transfer and kinematics that lead to ionization being more favorable than excitation. Unlike plane wave projectiles, vortex electrons possess non-zero transverse momentum, which as discussed below, can alter the kinematics of the collision causing noticeable changes in the cross sections. However, for small opening angles, this transverse momentum is small, and the vortex cross sections thus resemble those of plane wave projectiles.

As opening angle increases, the maximum in the cross section shifts to lower energies and a strong enhancement is observed at energies below 135 eV. This enhancement is a direct result of the vortex projectile's non-zero transverse momentum. To understand this shift in the maximum, consider a plane wave projectile at a fixed energy. The total excitation cross section results from an integration of the angle-differential cross section over scattering angle. This angle-differential cross section is sharply peaked at a particular scattering angle, corresponding to a narrow range of preferred momentum transfer values. Momentum transfer is a vector, defined as the difference between the incident and scattered projectile momentum ($\vec{q} = \vec{k}_i - \vec{k}_f$). For plane waves, both the direction and magnitude of $\vec{q}$ are well-defined. Thus, because the total excitation cross section at a given energy results from a sharply peaked angle-differential cross section corresponding to a specific momentum transfer, it is possible to correlate the peak of the total cross section with a specific momentum transfer unique to that projectile energy. In the case of hydrogen excitation to the n = 10 circular Rydberg state, the plane wave cross section ($\alpha = 0$), has its peak near 150 eV. The associated differential cross section has a sharp peak near $\theta = 14°$, which corresponds to a momentum transfer magnitude of $q = 0.8$ a.u. Thus, excitation to the n = 10 circular Rydberg state is most favorable for a momentum transfer of 0.8 a.u. Because vortex projectiles have a non-zero transverse momentum, this same momentum transfer magnitude can be achieved with a lower projectile energy. Thus, the peak in the total excitation cross section shifts to lower energy. As the opening angle of the twisted electron increases, so too does its transverse momentum, leading to a larger shift in the peak of the total cross section. Because this shift results from the kinematics of the projectile, it is independent of the target and is prominently observed in the total cross sections of Fig. 1 for all target atoms.

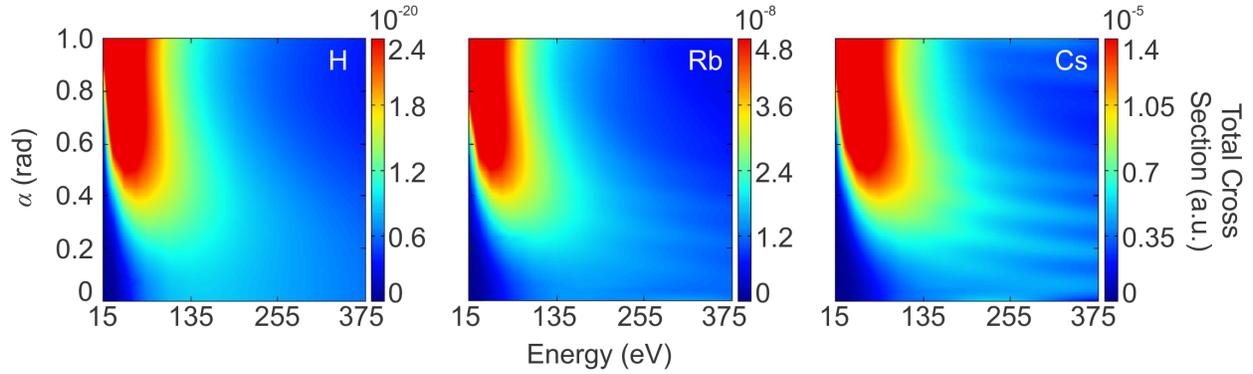

Figure 1 Total cross sections as a function of projectile energy and opening angle $\alpha$ for twisted electron excitation of macroscopic hydrogen, rubidium, and cesium targets from their respective ground states (1s, 5s, and 6s) to circular Rydberg states (n = 10 hydrogen; n = 14 rubidium and cesium). The maximum color value was set to the maximum cross section at $\alpha$ = 0.5 rad for each atom.

In addition to the transverse momentum of the twisted electron altering the magnitude of the momentum transfer, it also alters its direction. In particular, because the twisted electron's incident momentum lies on a cone, the momentum transfer vector also lies on a cone and no singular direction for $\vec{q}$ can be identified [38]. Nonetheless, it is possible to assess the role of the twisted electron's transverse momentum on the momentum transfer and also the excitation cross sections. Figure 2 shows how the energy required to yield a specific magnitude of momentum transfer varies with opening angle. Each curve corresponds to a fixed momentum transfer magnitude. Results are shown for an incident momentum azimuthal angle of $\phi_{k_i} = \pi/2$ and scattering angle of 14° (other selections of $\phi_{k_i}$ and scattering angle yielded qualitatively similar results). As the opening angle increases, less energy is required to maintain the same momentum transfer magnitude. Therefore, because the maximum excitation cross section favors a given momentum transfer magnitude, it is reasonable that this maximum shifts to smaller energy for twisted electrons with larger opening angles.

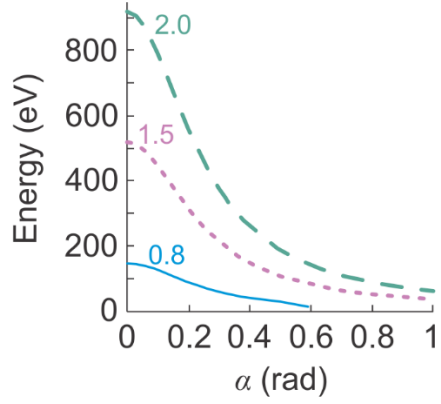

Figure 2 Energy required to achieve a given magnitude of momentum transfer as a function of twisted electron opening angle $\alpha$. Each curve has a fixed value of the momentum transfer magnitude (long dash, $q = 2$ a.u.; short dash, $q = 1.5$ a.u., solid $q = 0.8$ a.u.). Results are shown for excitation of hydrogen from the ground state to the $n = 10$ state with incident momentum azimuthal angle $\phi_{k_i} = \frac{\pi}{2}$ and scattering angle $\theta = 14°$.

While the shapes of the total excitation cross sections shown in Fig. 1 are similar for each of the atomic targets, the overall magnitude of the cross sections increases dramatically as the atomic number of the target increases. This increase can be traced to the lower excitation energy required to reach the rubidium or cesium circular Rydberg state. Additionally, finger-like structures are observed in the cross section for rubidium and cesium at large energies. These structures are likely due to interference effects between the transverse structure of the twisted electron density and the 5s or 6s radial wave functions of the rubidium and cesium ground states. Because the ground state of hydrogen is highly localized near the nucleus with no nodal structure, these interference effects are not present in the hydrogen cross sections.

## B. On-axis collisions provide clues to OAM contributions

While the cross sections for a macroscopic target do not explicitly depend on the OAM of the twisted electron, in the case of hydrogen, insight into the role of the OAM in these collisions can be achieved by examining on-axis ($\vec{b} = 0$) collisions with a single atom target. For hydrogen, the target electron density is localized near the nucleus, and therefore an on-axis collision has a large overlap between the projectile and atomic electron densities. Such a collision is thus expected to provide a dominant contribution in the average over impact parameters for the macroscopic target. Figure 3 shows the total cross sections for excitation of hydrogen from the ground state to the n = 10 circular Rydberg state as a function of projectile energy and opening angle for on-axis collisions. Calculations are presented for a range of OAM values.

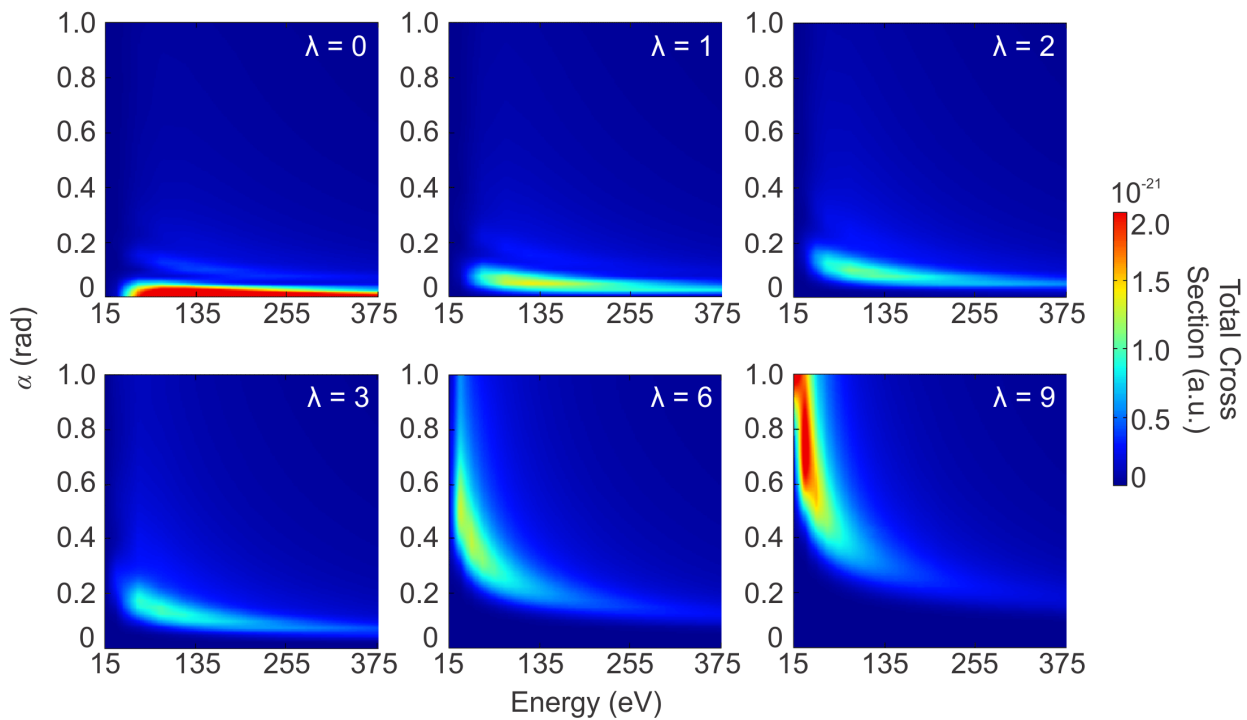

Figure 3 Total excitation cross sections as a function of projectile energy and opening angle $\alpha$ for an on-axis ($\vec{b} = 0$) collision with ground state hydrogen. The final state is the n = 10 circular Rydberg state. The maximum color value was set to half of the maximum of all cross sections, which occurred for a projectile with $\lambda = 0$.

At low values of OAM ($\lambda \leq 2$), the cross sections are largest at small opening angles and show a broad distribution across energy. This is expected because for small values of $\lambda$ and small opening angles, the momentum transfer remains small and is near optimal to cause excitation. As opening angle increases, the projectile's transverse momentum increases causing an increase in the momentum transfer and subsequent suppression of the cross section [39]. In contrast, for large values of OAM, the cross sections at small opening angles are suppressed. This is a result of the nodal structure of the projectile's radial density. For $\lambda > 0$, there is a node at the center of the projectile density on the propagation axis. This node increases in width as OAM increases, and thus for on-axis collisions with large OAM, there is very little overlap between the projectile and target electron densities. As opening angle increases, the cross sections for larger values of OAM increase and become well-localized at low energy. This is a result of two complimentary effects. First, larger opening angles reduce the width of the node in the projectile density, leading to increased overlap with the target electron density. Second, the projectile's larger transverse momentum yields a favored momentum transfer value at lower energies, making excitation more likely in this regime. Combined, these two effects lead to the enhanced cross sections at low energy and large opening angle when the projectile's OAM is large.

While the cross sections for a macroscopic target found by averaging over impact parameter are not strictly the same as an average of the cross sections over OAM, the on-axis cross sections of Fig. 3 point to the role of OAM in the cross sections for a macroscopic hydrogen target. The broad cross section structure as a function of energy that was observed at small opening angles for the macroscopic target likely results from the low OAM contributions, while the strongly peaked cross section at large opening angle and small energy is likely to arise from large OAM contributions.

## 4. Conclusions

Circular Rydberg states have emerged as a leading technology for quantum computing and quantum simulations, offering long lifetimes and strong interactions. As their use and application has grown, so too has the search for more efficient methods of generation. We presented total excitation cross sections for the generation of circular Rydberg states from the ground state of hydrogen, rubidium, and cesium atoms using twisted electrons. The unique features of twisted electrons, including their quantized OAM and non-zero transverse momentum, lend themselves to enhanced cross sections for excitation to high angular momentum atomic states. We showed that the excitation cross sections are enhanced for projectiles with large opening angles and low energies, and that in the case of hydrogen, this enhancement can be traced to contributions from projectiles with high values of OAM. Our results offer an alternative technique for the creation of circular Rydberg states that does not require multiphoton processes and is not subject to dipole selection rules, providing an additional feasible pathway for generating highly excited atomic states with large OAM.

## Acknowledgements

We gratefully acknowledge the support of the National Science Foundation under Grant No. PHY-2207209.